\documentclass[12pt]{iopart}
\usepackage{iopams}  
\usepackage{graphicx}  
\usepackage{amsfonts}  
\usepackage{mathrsfs}  

\newcommand{\mb}[1]{{\rm #1}}

\newcommand{\ket}[1]{\, | #1 \rangle}
\newcommand{\ii}{\;\! \mbox{i} \;\!}
\newcommand{\iis}{\;\! \mbox{\scriptsize i} \;\!}

\newcommand{\bbN}{\mathbb{N}}

\newcommand{\ps}{\ps}

\newcommand{\Hop}{{\hat H}}

\newcommand{\Up}{{\hat{\mathcal U}}}

\newcommand{\Np}{{\hat{\mathcal N}}}

\newcommand{\Ub}{\Up_{\beta}}

\newcommand{\phira}{{\phi_{\mbox{\tiny RA}}}}

\begin{document} 

\title[Saturation of Fidelity in the Atom-Optics Kicked Rotor]
{Saturation of Fidelity in the Atom-Optics Kicked Rotor}

\author{Sandro Wimberger\dag \, and  Andreas Buchleitner\ddag}

\address{\dag Dipartimento di Fisica, 
              Universit\`a di Pisa, Largo Pontecorvo 3, I-56127 Pisa}

\address{\ddag  Max-Planck-Institut f\"ur Physik komplexer Systeme, 
           N\"othnitzer Str. 38, D-01187 Dresden}

\begin{abstract}

We show that the quantum fidelity is accessible to cold atom experiments
for a large class of evolutions in periodical potentials, 
properly taking into account the experimental
initial conditions of the atomic ensemble. We prove 
analytically that, at the fundamental quantum resonances of the Atom-Optics
Kicked Rotor, the fidelity saturates at a constant, time-independent
value after a small number of kicks. The latter saturation arises
from the bulk of the atomic ensemble, whilst for the resonantly 
accelerated atoms the fidelity is predicted to decay slowly according to a 
power law.

\end{abstract}



\section{Introduction}


One of the most intriguing quantum phenomena
is the coherent evolution and control of a superposition of state vectors, 
which is at the heart of quantum parallelism in quantum computation schemes. 
One way to directly probe coherent quantum evolution is offered by 
the fidelity, which correlates the time evolution of a reference 
system with its slightly perturbed dynamics through  
$F(t) = \left|\langle \psi_0 |e^{\iis \hat{H}_{\Delta}t/\hbar}
e^{-\iis \hat{H} t/\hbar}| \psi_0 \rangle \right|^2$.
Here, ${\hat H}$ and ${\hat H}_{\Delta}={\hat H}+
\Delta {\hat V}$ represent the reference Hamiltonian and its perturbed 
variant, respectively, which generate the evolution, starting from 
the same initial state $|\psi_0 \rangle$. $F(t)$ was introduced as a measure
for the stability of quantum motion with respect to changes in some control
parameter $\Delta$ \cite{Per84}. 
The fidelity has recently evoked
considerable interest as an alternative route for the study of the effects of
perturbations on the coherent evolution of quantum systems, 
in particular in the context of quantum-classical
correspondence \cite{fidelity} and of quantum information processing
\cite{NC2001}. As for the former, the finite size of $\hbar$ makes the
classical definition of chaos through the exponential divergence of initially
nearby trajectories meaningless on the quantum level. However, the sensitivity
to changes in $\Delta$ prevails both in the
classical as well as in the quantum realm, and is dynamically quantified by
$F(t)$. 


Whilst previous studies focused on the limit of small
effective Planck constant \cite{fidelity,prosen}, we present in this Letter
an analytic result which is valid (a) in the {\em deep quantum regime}, and
(b) for {\em arbitrary} values of the control parameter $\Delta$. We study the
fidelity for cold, periodically kicked atoms, which mimic a paradigmatic 
model of quantum chaos theory, the $\delta$-Kicked Rotor (KR) \cite{Izr90}. 
In this context, we show that bona fide fidelity is readily accessible,
yet has never been measured in an experiment such as
reported in \cite{schlunk}. Specifically, we derive an analytic 
expression for the 
fidelity at the fundamental quantum resonances of the KR. 
At these resonances, a subclass of the experimental atomic ensemble 
\cite{kr,QR,wilson} experiences an enhanced acceleration by the kicking 
field, at precisely defined kicking periods \cite{Izr90}. We show that this
resonantly driven subclass induces a slow algebraic decay of fidelity,
whilst the bulk of the initial atomic ensemble produces the novel phenomenon
of a perfect saturation of fidelity at a finite and experimentally
easily detectable level. 

Quantum resonances are of purely quantum origin, although it has been
recently shown that they can be mapped onto a pseudo-classical problem. 
In this sense, we study
the quantum fidelity in the deep quantum regime, far from the semiclassical
limit of large actions, but close to
the {\em pseudo-classical} limit.
The latter anchors the near-resonant quantum dynamics to 
elliptic, stable resonance islands in the phase space 
of a suitably defined classical map,
precisely alike the usual semiclassical
limit. However, the pseudoclassical limit is reached at fixed and finite
$\hbar$ \cite{WGF}. Notwithstanding this formal complication,  
the correspondence between (pseudo)classically stable dynamics and
near-resonant quantum evolution immediately suggests
the remarkable robustness of the quantum 
dynamics at resonance, originally believed to be very sensitive to
perturbations (see \cite{Wim2004} for a further discussion). Hence, quantum
resonances provide a novel and {\em robust} tool of quantum control, and
fall within a larger class of particularly robust modes originating from the
dynamics of strongly perturbed quantum systems \cite{WGF,nonlin}. 
This robustness manifests indeed in the temporal evolution of fidelity
as shown in the sequel.

\section{The Atom-Optics Kicked Rotor and Quantum Resonant Motion}

The experimental realization \cite{kr,QR} of the KR uses 
cold atoms periodically kicked by an optical lattice. 
The Hamiltonian that generates the evolution of the atomic wave function
in its centre-of-mass degree of freedom
is given in the following dimensionless form \cite{zoller}
\begin{equation}
\Hop (t) =\frac{\hat{P}^2}{2} + k\cos(X)\sum_{m=-\infty}^{+\infty}
 \delta (t-m\tau)\;,
\label{ham}
\end{equation}
where $\hat{P}$ is the momentum operator, and 
$X$ the atom's position along the pulsed optical lattice. The 
two parameters that control the dynamics are the kicking period $\tau$, which
takes the role of an effective Planck constant in our units \cite{Izr90}, 
and the kick strength $k$.
The Hamiltonian  (\ref{ham}) describes the Atom-Optics Kicked Rotor (AOKR)
with the atoms moving along a line rather than on a circle (as in the
standard KR model). (\ref{ham}) induces the
characteristic properties of the quantum KR, which
sensitively depend on the parameter $\tau$ \cite{Izr90,WGF}. 
If $\tau=4\pi r/q$, with $r,q$
integers, then the kicking period is rationally related to the propagation
time of the kicked atoms across the lattice constant (which is $2\pi$ in our
units), and quantum transport is enhanced. Otherwise classically 
diffusive energy growth is suppressed by dynamical localisation 
\cite{Izr90}. 
In the following, we derive  analytical results for the fidelity, for the
experimentally relevant quantum resonances at 
$\tau=2\pi\ell$ ($\ell \in \bbN$) \cite{QR}, {\em i.e.}, 
for {\em large} effective Planck constants $\tau$.

An atom subject to a potential periodic in space and time 
is described by a wave packet composed of $2\pi$-periodic Bloch states,
$\psi_\beta(x+2\pi)=\psi_\beta(x)$, with the quasimomentum $\beta$. 
In our units, the latter is given by the fractional part of the momentum
$p=n+\beta, n \in \bbN$. 
$\beta$ is a constant of motion, so the different Bloch states 
evolve independently of one another, and their momenta
only change by integers. At $\tau=2\pi\ell$, 
a special behaviour occurs only for a specific, small {\em subclass}
of quasimomenta $\beta$. In much the same way as spatial periodicity enforces
ballistic motion in physical space, the one-period evolution 
operator is periodic in momentum space, enforcing ballistic propagation of the
corresponding $\beta$-states in momentum space. As shown in \cite{WGF}, 
during one period $\tau$, $\ket{\psi_\beta}$ evolves into 
$\Ub \ket{\psi_\beta}$, with 
\begin{equation} 
\label{ubeta} 
\Ub \;=\;e^{-\iis k\cos(\theta )}\;e^{-\iis\frac{\tau}{2}(\Np+\beta)^2}, 
\end{equation} 
where $\theta=X{\rm mod}(2\pi)$, and 
$\Np=-\ii d/d\theta$ is the angular momentum operator.
For instance, at $\tau=4\pi$, and $\beta=0,1/2$, or at $\tau=2\pi$ and
$\beta=1/2$, the second factor of $\Ub$ equals the identity
(apart from a global phase for $\beta \neq 0$), 
and the energy of the corresponding $\beta$-state grows ballistically, 
{\em i.e.}, quadratically with the number $N$ of kicks \cite{Izr90}. 
For general values of $\beta$, but $\tau = 2\pi \ell$, $\ell$ integer,
the state after the $N$th kick is, in the
momentum representation \cite{WGF}
\begin{equation} 
\label{ubeta1} \hspace{-2cm}
 \langle n \ket {\Ub^N {\psi _{\beta}}} = 
 e^{\iis n\mbox{\rm\scriptsize arg}(W_N)} 
 \int_0^{2\pi} \frac{d\theta}{\sqrt{2\pi}}e^{-\iis n\theta - \iis k 
 |W_N|\cos(\theta)} \psi_{\beta}
 \left(\theta-N\xi-\mbox{\rm arg}(W_N)\right)\;,
\end{equation} 
where $W_N=W_N(\xi):=\sum_{s=0}^{N-1}\exp(-\ii s \xi )$. 
If the initial state of the atom is a plane wave 
of momentum $p_0=n_0+\beta$, then $\xi$ takes the constant value 
$\xi_0 = \pi \ell (2\beta-1)$. If $\xi_0\neq 0$, then
$|W_N| =|{\sin(N\xi_0/2)}/{\sin(\xi_0/2)}|$,
else if $\xi_0=0$ then $|W_N|=N$, and 
the average kinetic energy  increases like $k^2N^2/4$ \cite{Izr90}.
The latter occurs at the resonant values 
$\beta=1/2+n/\ell$ mod$(1)$, $n=0,1,..,\ell-1$, 
which support {\em ballistic} motion \cite{WGF}. 
The remaining Bloch components of the original wave packet, with $\beta$
not belonging to the resonant subclass, undergo a quasiperiodic energy
exchange with the kicking field, leading to a {\em finite} 
spread of the associated momentum distribution for all times \cite{WGF}. 
How does a perturbation in the kick strength $k$ affect such 
quantum dynamics at quantum
resonance -- brought about by the abovementioned complete phase
revival in-between two successive kicks? The answer to this question will be
given by computing the fidelity of the resonant time evolution, using
eq.~(\ref{ubeta1}) for two different values of $k$. 
Before doing so, we show 
how to extract the fidelity from observables directly measurable in the
experiment \cite{schlunk}. 

\section{Experimental Measurement of Fidelity}

\begin{figure}
\centering \includegraphics[height=6cm]{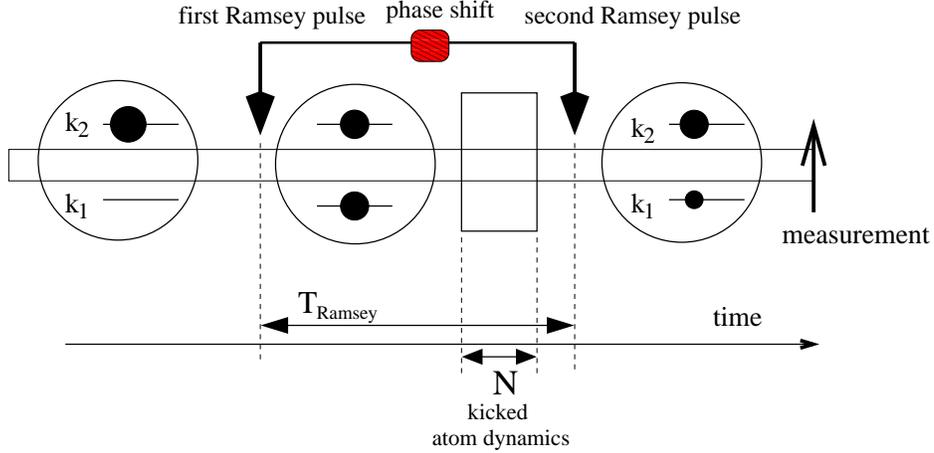} 
\caption{Experimental setup of \protect\cite{schlunk}. 
Each atom is prepared in the upper state (left), then a $\pi/2$ 
Ramsey pulse is applied. The dynamics of the coherent superposition 
of the upper (where $k=k_2$) and lower ($k=k_1$) level lasts for 
$N$ kicks, before a second $\pi/2$
Ramsey pulse is applied (with time delay  $T_{\mbox{\tiny RA}}$ with respect 
to the first one). In principal, any type of dynamics may occur in-between the
two Ramsey pulses, but our analytical results restrict here to the AOKR, as
an example for a coherent atomic evolution in a periodical optical lattice.
}
\label{fig:ramsey} 
\end{figure}

The experiment uses two sublevels of the atomic ground state of the
electronic degree of freedom, which couple differently
to the kick potential. Before the kicks are applied, the sublevels are
equally populated, and then evolve coherently. 
The splitting of the population is performed by a
Ramsey-type interferometer technique \cite{schlunk,AKD2003}, 
and the complete time evolution 
of the atoms is sketched in Fig.~\ref{fig:ramsey}. 
The internal states are represented by $(\psi_\beta^{(1)},
\psi_\beta^{{(2)}})$, where the upper index denotes the 
corresponding electronic level of the atom. 
$E_1,E_2$ are the energies of the two levels.
The Ramsey pulses are tuned to the electronic transition between 
these two levels, {\em i.e.}, they have a frequency 
$\omega_{\mb{\tiny RA}} \simeq (E_2-E_1)/\hbar$.
For simplicity, we assume instantaneous Ramsey pulses which occur immediately
before the first and after the last kick, respectively. These pulses couple 
the two states whilst they evolve coherently, but independently,
under the kicking potential. Let $\Ub^{(1)}$ and $\Ub^{(2)}$ be 
the propagators (\ref{ubeta}) in the upper 
and in the lower atomic level, respectively, and $\phi$ the relative phase of
the two Ramsey pulses. 
For an initial state where all the 
population resides in one of the electronic sublevels, {\em i.e.}, 
$\psi_\beta^{(1)}=0\;,\psi_\beta^{(2)}=\psi_\beta$,
the states after the final Ramsey pulse are \cite{Wim2004}
\begin{eqnarray*}
\psi_\beta^{(1)}(N,\phi)&=&
\frac12 e^{-\iis E_1N}
\left(\left( \Ub^{(1)}\right)^N e^{-\iis \phira} + 
\left( \Ub^{(2)}\right)^N \right)\psi_\beta
\nonumber\\
\psi_\beta^{(2)}(N,\phi)&=&\frac12 e^{-\iis E_2N}
\left(-\left( \Ub^{(1)}\right)^Ne^{\iis \phira}+
\left( \Ub^{(2)}\right)^N\right)\psi_\beta\;,
\end{eqnarray*}
with the phase $\phira \equiv (E_2-E_1)N -\phi$.
Projection onto a momentum eigenstate with momentum $n$ 
gives the momentum distribution, and we obtain 
the total probability in state $\psi_\beta^{(1)}$ upon summation over all $n$:
\begin{equation}
P_1(N,\phira)= \frac12\left[ 1+\mbox{\rm Re}\left\{ e^{-\iis \phira}
\langle \left( \Ub^{(2)}\right)^N\psi_\beta|
\left( \Ub^{(1)}\right)^N\psi_\beta\rangle\right\}\right] \; .
\label{eq:1}
\end{equation}
The observable given by eq.~(\ref{eq:1}) can be accessed experimentally
by measuring the atomic momentum distribution. 
The measurement, however, provides only an {\em average 
over all} $\beta$ in the initial ensemble \cite{schlunk}.
Thus, when the initial pure state $\psi_\beta$ (of the atomic centre-of-mass
motion) is replaced by a mixture which is described by the 
statistical density operator 
$
{\hat\rho}=\int_{0}^{1} d\beta\;
f(\beta ) |\psi_{\beta}\rangle\langle\psi_{\beta}|,
$
the final, measurable population in state ${\psi_\beta^{(1)}}$ is
\begin{equation}
{\overline P}_1(N,\phira)=\frac12\left\{1 +  \sqrt{F(N)}\cos(\phi'')
 \right\}\;.
\label{eq:2}
\end{equation}
We defined \cite{Wim2004}
\begin{equation} \hspace{-1cm}
\sqrt{F(N)}\cos(\phi'') \equiv \int_0^1 d\beta 
f(\beta )\mbox{\rm Re}\left\{e^{-\iis \phira}
\langle \left( \Ub^{(2)}\right)^N\psi_{\beta}|
\left( \Ub^{(1)}\right)^N\psi_{\beta}\rangle\right\} \;,
\label{eq:3}
\end{equation}
where, at fixed time $N$, $\phi''$ differs from $\phira$, as well as from 
$\phi$ by a constant shift, and
\begin{equation}
F(N)=\left| \int_{0}^{1} d\beta\;
f (\beta ) \langle\left( \Ub^{(1)}\right)^N\psi_{\beta}|
\left( \Ub^{(2)}\right)^N\psi_{\beta}\rangle\right|^2 
\label{eq:fidy_exp}
\end{equation}
corresponds to the fidelity for an incoherent ensemble of
$\beta$-states. From this analysis, we see that the 
fidelity is equal to the squared 
difference between the maximum and the minimum values 
taken by  ${\overline P}_1(N,\phi)$, 
while the original phase $\phi$ varies in 
$[0,2\pi)$ \cite{Wim2004,HBSSR2005}. 

Hence, the fidelity is accessible as the visibility of the integrated momentum
distribution (\ref{eq:2}) as a function of the Ramsey phase $\phi$
by the experimental setup of Fig.~\ref{fig:ramsey}. One must, however,
explicitly take into account the scan of the Ramsey phase,
in contrast to what has been dubbed ``fidelity'' in \cite{schlunk}.
Moreover, as long as the evolution operators are known and the atoms
evolve in a spatially periodical potential, eqs.~(\ref{eq:1}-\ref{eq:fidy_exp})
remain valid and are not restricted to the here analysed AOKR.

\section{Analytical Results and Discussions}

For the AOKR experiments \cite{kr,QR}, 
with an initially Gaussian momentum distribution with root mean square
$>1$, the distribution of quasimomenta 
$f_0(\beta)$ is nearly uniform \cite{WGF}. Since $\beta$ is a constant of
motion, the fidelity is
$$
F(N) \equiv \left | 
\int d\beta f_0(\beta) \sum_{n=-\infty}^{+\infty}  \psi^{(1)*}_\beta (N,n) 
\psi^{(2)}_\beta (N,n) \right |^{2}\;.
$$
To compute $F(N)$ we must calculate the overlap 
of two wave functions (\ref{ubeta1}) for two values of $k$, 
and sum over all integer momenta $n$.
For the case when the state of the atom is a plane wave with  
momentum $p_0=n_0 + \beta$, we have $|\psi_{\beta}(\theta-N\xi-
\mbox{\rm arg}(W_N))|^2=1/(2\pi)$, leading to the 
result for the fidelity for {\em one} $\beta$-state:
$ F_{\beta}(N) = J_0^2\left( | W_N|(k_1-k_2)\right)$.
For $\Delta k \equiv k_1-k_2 =0$, $F_{\beta}=F_{\beta}(N)=1$.
For resonant $\beta$ we have $|W_N|=N$, and we can use the
asymptotic expansion formula $[9.2.1]$ from \cite{AS72} for the 0th order
Bessel function $J_0$.
The fidelity for large times at quantum resonance is then
\begin{equation}
\label{overfid1}
F_{\beta_{\mbox{\tiny res}}}(N) \simeq \frac{2}{\pi \Delta k N} \cos^2 
\left(\Delta k N -\frac{\pi}{4}\right) \;.
\end{equation}
This shows that the fidelity falls off as a power law 
$\propto 1/N$ for the resonant $\beta$-states. 
Another particular case is given for $\beta =0$:
$F _{\beta=0}(N) = J_0^2 \left( |\sin (N\pi/2)|\Delta k \right)$,
which at $N=1$ depends just on $\Delta k$.
Then the fidelity can be made arbitrarily 
small when $\Delta k \simeq z_0$, where 
$z_0 \simeq 2.40$ is the first zero of $J_0$. In turn, the fidelity
can be maximised by maximising $J_0^2$, what provides an option
for coherent control when quasimomentum can be precisely controlled
in experiments with Bose-Einstein condensates \cite{wilson}.

\begin{figure}
\centering  \includegraphics[height=11.5cm,angle=270]{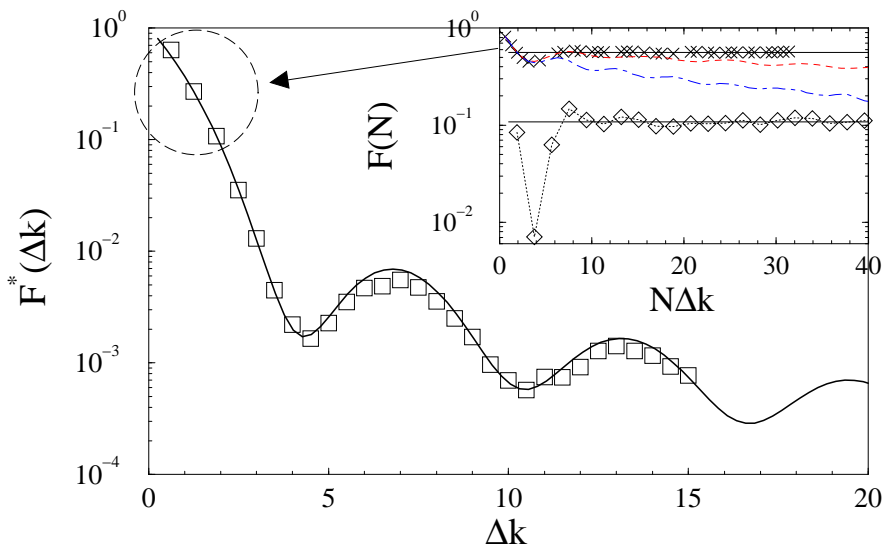} 
\caption{
Asymptotical formula for the fidelity for $N \to \infty$ 
from eq.~(\ref{ftfinal}) (solid line),
compared with numerical data (squares) 
obtained by evolving ensembles of $10^4$ 
$\beta-$rotors with an initially uniform momentum distribution in $[0,1)$, at
$\tau=2\pi$ and $N=50$. Inset: 
numerically computed fidelity vs. $N\Delta k$ for $\tau=2\pi$, 
$k_1=0.8\pi$, and fixed $\Delta k=0.6283$ (crosses), $1.885$ (diamonds).
The position of the minima corresponds to the time 
$3.83\tau /\Delta k$. The fidelity saturates for 
times $N \gtrsim 20$ at $\Delta k$-dependent, constant values, indicated by the
horizontal lines. Data for finite detunings $\epsilon = 0.025$ (dashed)
and $0.1$ (dot-dashed) is shown for $\Delta k=0.6283$.
}
\label{fig:fidyasym}
\end{figure}

To compute $F(N)$ for $f_0(\beta) \equiv 1$, what approximates very
well the experimental situation in \cite{QR,WGF},
we perform the average over the different quasimomenta, i.e.
\begin{eqnarray}
 &&\int_0^1 \; d\beta  J_0\left(|W_N (\xi)| \Delta k \right)
 = \int_{-\pi}^{\pi}\frac{dx}{2\pi}J_0
 \left( \Delta k\sin(Nx)\csc(x)\right)
 \nonumber\\
 &=& \int_{-\pi}^{\pi}\frac{dx}{4\pi^2}
 \sum\limits_{r=0}^{N-1}\frac{2\pi}{N}J_0
 \left( \Delta k \sin(x)\csc(xN^{-1}+2\pi rN^{-1}) \right)\;. 
\label{eq:fourcg}
\end{eqnarray}
We used the definition of $W_N(\xi)$ as given after
eq.~(\ref{ubeta1}), and the periodicity of the sine in the argument 
of the Bessel function. In the limit when $N\to\infty$ and  
$2\pi r/N\to\alpha$, the sum over $r$ approximates the integral 
over $\alpha$, and (\ref{eq:fourcg}) converges to 
\begin{equation*}
  \int_0^1 \; d\beta J_0 \left( |W_N(\xi)| \Delta k \right)
 \rightarrow \frac{1}{(2\pi)^2} \int_{-\pi}^{\pi}dx \int_0^{2\pi}
 d\alpha\;J_0 \left(\Delta k \sin(x)\csc(\alpha) \right)\;.
\end{equation*}
With the help of $[11.4.7]$ from \cite{AS72}, this leads 
us to the final result for the asymptotic value $F^*$ of the fidelity 
\cite{Wim2004}
\begin{equation} 
\label{ftfinal} 
F^*(\Delta k) \equiv \frac{1}{(2\pi)^2} \left( \int_0^{2\pi} 
d\alpha\;J^2_0 \left( \frac{\Delta k \csc(\alpha)}{2} \right) \right) ^2\;. 
\end{equation}
The dependence on $\Delta k$ of the formula (\ref{ftfinal}) is shown in 
Fig.~\ref{fig:fidyasym}, together with numerical computations.
$F^*(\Delta k)$ oscillates quasiperiodically  rather than to drop monotonously 
with $\Delta k$.
Note that the fidelity always saturates at exact 
resonance $\tau=2\pi \ell$ ($\ell
\in \bbN$), and the saturation time and level depend on 
$\Delta k$, as illustrated in the inset. We interprete our analytical
result (\ref{ftfinal}) by referring to the wave 
function in eq.~(\ref{ubeta1}): the perturbation enters in the expression for
the fidelity in the form of phase factors 
$\exp\left(-\ii \Delta k |W_N|\cos(\theta)\right)$. For non-resonant 
$\beta$, $|W_N|\to 0$ in the limit $N\to \infty$, and hence 
these phases eventually tend to unity, {\em i.e.}, 
the perturbation completely looses its effect on the relative 
evolution of the two wave packets under comparison. Note that a 
fidelity plateau, somewhat reminiscent of the saturation observed here, 
was recently predicted for some model systems
\cite{prosen}. However, this strictly semiclassical 
result was obtained in the perturbative limit,
and the plateau observed in \cite{prosen}
only represents a transient effect, followed by the subsequent
decay of fidelity. By contrast, our formula (\ref{ftfinal}) is a result valid
for all (sufficiently large) $N$. 

The asymptotic saturation value $F^*$ depends on the
difference $\Delta k$ in the kick strength and, in addition, on the
distribution of the initial conditions $f_0(\beta)$: 
the resonant $\beta$ -- which 
behave according to (\ref{overfid1}) -- determine the initial
decay and the consecutive rise of the fidelity $F(N)$ shown in the inset 
of Fig.~\ref{fig:fidyasym}.
Their contribution quickly decreases, since the speed
of the wings of the two wave packets to be compared is proportional
to the two different $k$ \cite{Izr90,WGF}. 
After this initial stage, the bulk of
non-resonant $\beta$ takes over and dominates then the fidelity, 
leading to rapid saturation.
These competing effects give rise to the damped oscillations around the
asymptotic value of fidelity. At fixed $\Delta k$, $F^*$ is the 
lower, the more weight is given to quasimomenta which are
close to resonant values. The initial decay of the fidelity is
due to nearly resonant $\beta$ values:
the minima in the inset in Fig.~\ref{fig:fidyasym} are found
by differentiating the time-dependent fidelity with respect to $\Delta k$.
Then, the first zero of the 1st order
Bessel function $J_1(z)$ ($z=z_1 \simeq 3.83$)
is responsible for the observed minima. 
Figure \ref{fig:fidyasym} also shows that 
strict saturation is destroyed by arbitrarily small detunings 
$\epsilon \equiv 2\pi \ell - \tau$ from resonance 
(since then the quasiperiodic motion for the non-resonant 
$\beta$ is destroyed). 
However, the temporal asymptotic decay of the
fidelity depends continuously on $\epsilon$, and 
is very slow for small $\epsilon$,
what leaves the here predicted saturation an 
{\em experimentally robust} observable (since
the kicking period is the best controlled parameter in the experiment 
\cite{schlunk,kr,QR}).

\section{Conclusions}

In summary, we analytically derived the asymptotic behaviour of fidelity for
kicked cold atoms at quantum resonance, far from the classical limit and
for arbitrary perturbations in the kicking strength $k$. 
The predicted saturation sets in after a small number of kicks, at an 
experimentally observable level. 
Note, in particular, that $F^*\geq 50\ \%$ for $\Delta k\sim O(k)$, what 
clearly stresses the robustness of quantum resonant motion.
The type of dynamics occurring in-between the two interferometric pulses
(see Fig.~\ref{fig:ramsey}) is not restricted to the AOKR. Our derivation
[see eqs. (\ref{eq:1}-\ref{eq:fidy_exp})] applies 
for arbitrary evolutions of cold atoms
in a periodical potential, since it properly takes into account the different
quasimomentum classes (corresponding to all the initial conditions) 
of the experimental ensemble.

\end{document}